\documentclass[10pt]{article}
\pdfoutput=1
\usepackage{jheppub}
\usepackage{enumerate}
\usepackage{bm}
\usepackage{bbm}
\usepackage{enumitem}
\usepackage[usenames,dvipsnames]{xcolor}
\usepackage{amsmath}
\usepackage{amsfonts}
\usepackage{amssymb}
\usepackage{graphicx}
\usepackage{hhline}
\usepackage{subfig}
\usepackage{hyperref}
\usepackage{verbatim}
\usepackage{amsmath,amsthm,amssymb}
\usepackage{lscape}
\usepackage{float}
\usepackage{caption}
\usepackage{lscape}
\usepackage{breqn}
\usepackage{multicol}
\usepackage{graphics}
\usepackage{tikz,todonotes}
\usepackage[utf8]{inputenc}
\usepackage{autobreak}
\usepackage{verbatim}
\usetikzlibrary{arrows,positioning,decorations.markings,decorations.pathmorphing,calc}
\pgfdeclarelayer{edgelayer}
\pgfdeclarelayer{nodelayer}
\usetikzlibrary{decorations.pathreplacing}
\pgfsetlayers{edgelayer,nodelayer,main}
\tikzset{none/.style={draw=none}}
\tikzset{new edge style 2/.style={black}}
\tikzset{new style 0/.style={black}}
\tikzset{rednode/.style={draw=none, scale=0.3pt,fill=red,circle, draw}}
\tikzset{redline/.style={line width=0.3mm,red}}
\tikzset{greyE/.style={line width=0.1mm,gray}}
\usepackage[utf8]{inputenc}
\usetikzlibrary{arrows,positioning,decorations.markings,decorations.pathmorphing,calc}

\definecolor{hyperref}{RGB}{026,028,087}

\newcommand{\beq}{\begin{equation}}
\newcommand{\eeq}{\end{equation}}
\newcommand{\bal}{\begin{aligned}}
\newcommand{\eal}{\end{aligned}}
\newcommand{\bea}{\begin{eqnarray}}
\newcommand{\eea}{\end{eqnarray}}
\def\be{\begin{equation}}
\def\ee{\end{equation}}

\def\beq{\begin{equation}}
\def\eeq{\end{equation}}

\newcommand{\K}{\mathcal K}
\newcommand{\Q}{\mathcal Q}

\renewcommand{\L}{\mathcal L}

\def\be{\begin{equation}}
\def\ee{\end{equation}}
\def\ba{\begin{eqnarray}}
\def\ea{\end{eqnarray}}

\def\nn{\nonumber}

\def\d{\mathrm{d}}

\usepackage[normalem]{ulem}

\def\ba{\begin{eqnarray}}
\def\ea{\end{eqnarray}}

\def\L{\mathcal{L}}
\def\K{\mathcal{K}}
\def\X{\mathcal{X}}
\def\Z{\mathcal{Z}}

\def\d{\mathrm{d}}
\def\mn{_{\mu \nu}}
\def\mnup{^{\mu \nu}}

\def\mupn{^\mu_{\phantom{\mu}\nu}}
\def\({\left(}
\def\){\right)}

\def\p{\partial}
\def\ie{{\em i.e. }}

\newcommand{\ud}[2]{^{#1}_{\phantom{#1}#2}}
\newcommand{\du}[2]{_{#1}^{\phantom{#1}#2}}

\begin{document}


\title{To Half--Be or Not To Be?}

\author[a,b]{Claudia de Rham,}
\author[c]{Sebastian Garcia-Saenz,}
\author[b,d,e]{Lavinia Heisenberg,}
\author[a]{Victor Pozsgay,}
\author[c]{and Xinmiao Wang}
\affiliation[a]{Theoretical Physics, Blackett Laboratory, Imperial College, London, SW7 2AZ, UK}
\affiliation[b]{Perimeter Institute for Theoretical Physics,
31 Caroline St N, Waterloo, Ontario, N2L 6B9, Canada}
\affiliation[c]{Department of Physics, Southern University of Science and Technology, Shenzhen 518055, China}
\affiliation[d]{Institute for Theoretical Physics, Heidelberg University, Philosophenweg 16, 69120 Heidelberg, Germany}
\affiliation[e]{Institute for Theoretical Physics,
ETH Z\"urich, Wolfgang-Pauli-Strasse 27, 8093, Z\"urich, Switzerland}

\emailAdd{c.de-rham@imperial.ac.uk}
\emailAdd{sgarciasaenz@sustech.edu.cn}
\emailAdd{lavinia.heisenberg@phys.ethz.ch}
\emailAdd{v.pozsgay19@imperial.ac.uk}
\emailAdd{12132943@mail.sustech.edu.cn}

\abstract{It has recently been argued that half degrees of freedom could emerge in Lorentz and parity invariant field theories, using a non-linear Proca field theory dubbed Proca-Nuevo as a specific example. We provide two proofs, using the Lagrangian and Hamiltonian pictures, that the theory possesses a pair of second class constraints, leaving $D-1$ degrees of freedom in $D$ spacetime dimensions, as befits a consistent Proca model. Our proofs are explicit and straightforward in two dimensions and we discuss how they generalize to an arbitrary number of dimensions. We also clarify why local Lorentz and parity invariant field theories cannot hold half degrees of freedom. }

\maketitle

\section{Introduction}
\label{sec:introConstraint}

The study of self-interacting massive spin-1 fields has received considerable interest in potential connection with dark energy, inflation and other cosmological and particle physics applications.
The generic ghost-free effective field theory (EFT) of a self-interacting massive spin-1 field includes the Generalized Proca (GP) interactions proposed in \cite{Tasinato:2014eka,Heisenberg:2014rta} (see \cite{Allys:2015sht,Allys:2016jaq,Jimenez:2016isa,Heisenberg:2018vsk} for related works). The key feature of GP is ensuring that the equations of motion of all degrees of freedom remain second order in derivative in the decoupling limit (DL), where the helicity-0 and -1 modes can be decoupled, ensuring the theory absence of Ostrogradsky-type ghosts in that limit. While the equations of motion generically include higher derivatives away from that limit, the presence of a non-trivial constraint ensures that the theory propagates the correct  $D-1$ number of degrees of freedom in $D$ dimensions.
These derivative self-interactions were shown to be responsible for genuinely new properties in relation to the screening mechanisms and the coupling to alternative theories of gravity \cite{DeFelice:2016cri,Heisenberg:2018acv,Garcia-Saenz:2021acj}, although the main motivation for the development of the theory was aimed at applications to astrophysical systems \cite{Chagoya:2016aar, Minamitsuji:2016ydr, Cisterna:2016nwq,Chagoya:2017fyl,Heisenberg:2017hwb, Minamitsuji:2017aan,Heisenberg:2017xda,Kase:2017egk,deFelice:2017paw,Nakamura:2017lsf,Kase:2018owh,Rahman:2018fgy,Kase:2018voo,Kase:2020yhw,Garcia-Saenz:2021uyv,Brihaye:2021qvc} and cosmology \cite{Jimenez:2013qsa,Emami:2016ldl,DeFelice:2016yws,DeFelice:2016uil,Heisenberg:2016wtr,Nakamura:2017dnf,Nakamura:2018oyy, Kase:2018nwt,Domenech:2018vqj,Oliveros:2019zkl,DeFelice:2020sdq,Minamitsuji:2020jvf,Heisenberg:2020xak,Garnica:2021fuu,Geng:2021jso,Oestreicher:2022zkh,Chagoya:2022wba}.

As indicated in \cite{deRham:2011rn,deRham:2011qq}, (see also \cite{deRham:2016wji}), when considering theories with multiple fields and particularly those involving gravity,  the relation between higher derivatives and Ostrogradsky-like instabilities is more subtle and more general classes of Beyond-second order GP theories were soon proposed in  \cite{Heisenberg:2016eld,Heisenberg:2016lux,Kimura:2016rzw,Seifert:2019xan,GallegoCadavid:2019zke}. Those  involve higher derivative equations of motion when coupled to gravity while maintaining the presence of a constraint that projects out the unwanted ghost.
The generalization to models with multiple spin-1 fields was considered in \cite{Allys:2016kbq,Jimenez:2016upj,Rodriguez:2017wkg,ErrastiDiez:2019trb, ErrastiDiez:2019ttn,Gomez:2019tbj,GallegoCadavid:2020dho,GallegoCadavid:2022uzn,Martinez:2022wsy,Gomez:2023wei,2301.05222}. Some of these non-linear Proca models have also been studied at the quantum level where they have been shown to describe consistent quantum field theories \cite{Amado:2016ugk,deRham:2017imi,deRham:2017xox,deRham:2018qqo,Ruf:2018vzq,ErrastiDiez:2021ykk}. See also \cite{Rodriguez:2017ckc} for a review.

As explained already, the correct number of degrees of freedom in GP is related to the fact that the equations of motion for the massless spin-0 and spin-1 modes of its DL are second-order in time derivatives and that the $A_0$ is constrained to be non-dynamical. However, there is no reason for this to be true in \textit{all} non-linear completions of the free Proca theory. Indeed, one could remove the additional ghostly degree of freedom by realising the constraint in a more complicated way, while the equations of motion need not be second-order if they are degenerate \cite{deRham:2016wji}.

In this vein, a genuinely new class of interactions was proposed recently in \cite{deRham:2020yet} under the name of Proca-Nuevo (PN). This class of theories was inspired by the vector sector of the DL of massive gravity \cite{deRham:2010kj,deRham:2010gu,deRham:2011rn,Ondo:2013wka}, a theory that has been shown to be ghost-free \cite{Hassan:2011hr,Hassan:2011tf,Hassan:2011ea,Bernard:2014bfa,Bernard:2015mkk}. The analysis performed in \cite{deRham:2020yet} proceeded to prove that, even though the equations of motion in the DL of PN were higher-order, the Hessian matrix of field velocities admitted a null eigenvector (NEV) which ensured the absence of any ghostly degree of freedom. It was subsequently noted that a subclass of GP interactions could safely be added to PN theories without spoiling the Hessian constraint, resulting in the Extended Proca-Nuevo (EPN) theory introduced in \cite{deRham:2021efp}. Importantly, it was also shown that GP and (E)PN theories are dynamically inequivalent by computing scattering amplitudes \cite{deRham:2020yet} and by providing different phenomenological predictions related to dark energy \cite{deRham:2021efp}. It was consequently shown that if one models dark energy with a time-dependent (E)PN vector condensate, the theory predicts a technically natural vector mass and dark energy scale \cite{Heisenberg:2020jtr,deRham:2021yhr}. Constraints on these theories by demanding that they should be embeddable within a standard high energy completion were explored in \cite{deRham:2017zjm,deRham:2018qqo,deRham:2022sdl}.\\

Let us turn now to the issue of the constraint algebra of vector field theories. Consider first electrodynamics. In this case, the $U(1)$ gauge symmetry ensures the existence of a first class constraint associated with the local symmetry,  removing a pair of conjugate variables in phase space, equivalently two degrees of freedom in field space, leading to $D-2$ propagating degrees of freedom in $D$ dimensions. The addition of a mass term (and more generally of non-$U(1)$-invariant self-interactions\footnote{The breaking of $U(1)$ invariance has to occur at the linear level about the vacuum to avoid infinitely strong coupling issues.}) breaks this symmetry and the first class constraint is downgraded to a pair of second class constraints, each removing half a degree of freedom (see \cite{Heisenberg:2014rta,BeltranJimenez:2019wrd,ErrastiDiez:2020dux} for constraint analyses of GP). The resulting theories then propagate $D-1$ physical degrees of freedom in $D$ dimensions. The claim of the absence of ghosts in (E)PN is based on the existence of a second class constraint realised in the form of a null equation for the Hessian matrix. The existence of a second class constraint removes one degree of freedom in the $2D$-dimensional Hamiltonian phase space, potentially leaving $D-1/2$ degrees of freedom in field space and not fully exorcising the Ostrogradsky ghost. It was then argued in \cite{deRham:2020yet,deRham:2021efp}, that there should exist a secondary constraint based on the fact that the existence of half degrees of freedom should not arise in Lorentz and parity invariant theories (see Section \ref{sec:HalfDegree}). Contrariwise, it was claimed in \cite{ErrastiDiez:2022qvd} that (E)PN might be the first counterexample to this expectation\footnote{A similar claim was made in \cite{Kluson:2011qe} that ghost-free massive gravity failed to have a secondary constraint. This constraint was later explicitly derived in \cite{Hassan:2011ea}. In the words of \cite{Hassan:2011ea}, ``{\it an odd dimensional phase space --- an odd situation indeed}''.}. Motivated by this, we present in this work a complete analysis of the constraint algebra of EPN, proving that (as expected), the primary second class constraint is followed by a secondary constraint which fully takes care of eradicating the whole would-be ghostly degree of freedom. \\

The rest of the paper is organized as follows: We start with a short discussion on the absence of half degrees of freedom in Lorentz and parity invariant theories in Section \ref{sec:HalfDegree}, followed by a review of (E)PN in Section \ref{sec:reviewEPNConstraint}, where we take particular care to explicitly derive the analytical form of the (E)PN Lagrangian, Hessian, NEV and Hessian constraint, focusing on the two-dimensional case. The constraint analysis is performed in Section \ref{sec:LagrConstraint} in the Lagrangian picture and in Section \ref{sec:HamConstraint} in the Hamiltonian picture. The extension to arbitrary spacetime dimensions is presented in Section \ref{sec:HigherDConstraint}, although here we limit our analysis to a minimal sub-class of (E)PN. In all cases, we establish the absence of propagating half degrees of freedom and thus the consistency of EPN theory.

\section{Half--Being}
\label{sec:HalfDegree}

Before diving into the subtleties related to interacting Proca theories and how the secondary constraint is realized, it is worth asking ourselves if and when half a number of dynamical field space degrees of freedom can propagate in a generic local field theory. We also refer to \cite{Golovnev:2022rui} for a useful review of the role played by constraints and degrees of freedom.

As explained in \cite{deRham:2014zqa}\footnote{See specifically the explanations below Eq.~(7.30) of
\href{https://link.springer.com/article/10.12942/lrr-2014-7}{Living Rev. Rel. 17 (2014) 7}.}, half a field space degree of freedom or an odd number of degrees of freedom in phase space corresponds to a system of first order differential equations that cannot be recombined into a fully second order system. This implies the existence of a {\it field space} degree of freedom (say $\chi$) whose dynamics (upon appropriate diagonalization) would be governed by a first order time derivative equation, $\p_t \chi=\hat{O}(\p_i) \chi+\cdots$, where ellipses involve other fields with no time derivatives acting on them and $\hat O$ is an operator that only involves functions of spacetime and spatial derivatives of any order but no additional time derivatives.

If the theory is fundamentally {\it Lorentz invariant} (even if considering a solution that spontaneously breaks Lorentz invariance), then this differential equation can necessarily be recast as $L^\mu \p_\mu \chi=f(x^\mu) \chi+\cdots$, for a Lorentz vector $L^\mu$ meaning that the evolution equations should contain terms that are linear in space derivatives, $L^i\p_i \chi$.  Under a parity transformation $x^i \to -x^i$, these terms are odd and the theory as a whole would therefore break parity. Alternatively, an odd number of dynamical phase variables can be propagating in theories that preserve parity at the price of breaking Lorentz invariance. Phrased in terms of the last odd phase space variable, locality joined together with Lorentz and parity invariance impose $L^\mu=0$ meaning that the equation for $\chi$ is none other than the additional constraint responsible for ensuring an even-dimensional phase space.

As a result, in complete generality, we can infer that a local, Lorentz and parity invariant field theory can never propagate an odd number of phase space degrees of freedom or equivalently half a number of field space degrees of freedom\footnote{A constraint-based proof of this fact was given in \cite{Crisostomi:2017aim} in the context of degenerate scalar field theories.}. An exception to this rule was suggested recently in \cite{ErrastiDiez:2022qvd}, which if correct would hint towards a potential undiagnosed loophole to the previous argument. However, as we shall see below, those conclusions were premature and upon appropriately identifying the constraint algebra the example suggested in \cite{ErrastiDiez:2022qvd} follows precisely the logic highlighted above. Rather than being a dynamical equation for half a field space degrees of freedom, the remaining equation is nothing other than a secondary second class constraint that projects out the other half field space variable, as expected from Lorentz and parity invariance. We shall see this more precisely in what follows.

\section{Review of (Extended) Proca-Nuevo}
\label{sec:reviewEPNConstraint}

PN interactions for a massive vector field were first introduced in \cite{deRham:2020yet} and further developed in its extended version (EPN) in \cite{deRham:2021efp} where Generalized Proca interactions were added to PN. (E)PN itself was inspired by the theory of massive gravity proposed in \cite{deRham:2010ik,deRham:2010kj,deRham:2011qq,deRham:2014zqa}, where its decoupling limit was shown to involve non-trivial vector interactions \cite{Ondo:2013wka,deRham:2016plk,deRham:2018svs} with a non-linearly realized constraint. The theory can be defined on any background geometry $g\mn$ so long as the mixing with the gravitational degrees of freedom remains under control; see \cite{deRham:2021efp} for details. In this work however, we consider for simplicity a flat $D$-dimensional flat background metric. The vector field is denoted by $A_{\mu}$ and its interaction energy scale $\Lambda$.
The formulation of the theory relies on the Lorentz tensor
\begin{equation}
	f\mn[A] = \p_\mu \phi^a \p_\nu \phi^b \eta_{ab}=\eta\mn + 2 \frac{\p_{(\mu} A_{\nu)}}{\Lambda^{D/2}} + \frac{\p_{\mu} A^\rho \p_{\nu} A_\rho}{\Lambda^D} \,,
\end{equation}
where
\begin{equation}
	\phi^a=x^a+\frac{1}{\Lambda^{D/2}}A^a\,.
\end{equation}
Even though $\phi^a$ is not a Lorentz vector, the object $f\mn$ is a Lorentz tensor. The (E)PN operators are built in terms of the symmetric polynomials of the tensor $\K\mupn$  defined as \cite{deRham:2010kj},
\begin{align}
	 \K\mupn &= \X\mupn -\delta \mupn \,, \\
	 \text{with }	\quad \X\mupn[A] &= \left( \sqrt{\eta^{-1}f[A]}  \right)\mupn \qquad \text{i.e.}\qquad \X^{\mu}_{\phantom{\mu} \alpha} \X^{\alpha}_{\phantom{\alpha} \nu} = \eta^{\mu \alpha}f_{\alpha \nu}[A]
\,. \label{eq:XX=f}
\end{align}
Mathematically there exists multiple branches of solutions satisfying the relation $\X \X =\eta^{-1}f$ \cite{Comelli:2015ksa}, but only solutions which are continuously connected to the trivial one for which $\X\mupn[0]=\delta\mupn$ should be considered.

With this choice in mind, the EPN theory for a massive vector field $A_{\mu}$ in arbitrary spacetime dimension $D$ is defined by the Lagrangian \cite{deRham:2020yet, deRham:2021efp}\footnote{Note that on flat spacetime, the $d_D$ term is a total derivative independently of the form of the arbitrary function of $X$ \cite{BeltranJimenez:2019wrd}.}
\begin{equation}
	\L_{\rm EPN}[A] = \Lambda^D \sum_{n=0}^D \alpha_n(X) \L_n[\K] + \Lambda^D \sum_{n=1}^{D-1} d_n(X) \frac{\L_n[\p A]}{\Lambda^{n D/2}} \,,
\label{eq:defLKconstraints}
\end{equation}
where the first sum contains the pure PN terms and the second involves the GP interactions introduced in \cite{Tasinato:2014eka,Heisenberg:2014rta}. 
The dimensionless generic functions $\alpha_n$ and $d_n$ depend on the Lorentz scalar $X$ defined as
\begin{equation}
	X = \frac{1}{\Lambda^{D-2}}\,A^{\mu}A_{\mu} \,,
\end{equation}
while $\L_n[\Q]$ stands for the $n$-th elementary symmetric polynomial of the eigenvalues of the matrix $\Q$,
\begin{equation}
	\L_n[\Q] = -\frac{1}{(D-n)!} \varepsilon^{\mu_1 \cdots \mu_n \mu_{n+1} \cdots \mu_D} \varepsilon_{\nu_1 \cdots \nu_n \mu_{n+1} \cdots \mu_D} \Q^{\nu_1}_{\phantom{\nu_1}\mu_1} \cdots \Q^{\nu_n}_{\phantom{\nu_n}\mu_n}\,.
\end{equation}
For instance, in any dimension, we have $\L_1[\Q]=[\Q]$ and $\L_2[\Q]=[\Q]^2-[\Q^2]$ (the Fierz-Pauli structure), where square brackets represent the trace of a matrix.

\subsection{Trivial vacuum}

We are only interested in theories for which we recover the standard free Proca theory perturbatively about the vacuum $\langle A_\mu \rangle=0$, so to quadratic order in the vector field, it is understood that the Lagrangian \eqref{eq:defLKconstraints} must reduce to
\ba
\L_{2}=-\frac 14 F\mn^2-\frac 12 m^2 A^2 + \mathcal{O}\(\frac{\p A^3, A^4}{\Lambda^{(D-4)/2}}\)\,.
\ea
In particular, as emphasized in \cite{deRham:2020yet}, the theory only makes sense if the helicity-0 mode of the vector field carries a kinetic term on the trivial  standard Lorentz invariant vacuum $\langle A_\mu \rangle=0$, which implies that the potential should always include a mass term. The functions $\alpha_n(X)$ and $d_n(X)$ should therefore be analytic functions of their argument about $X=0$, so we can express them in terms of their Taylor expansion
\ba
\label{eq:expansion_alpha_d}
\alpha_n(X)=\sum_{n\ge 0} \frac{1}{k!}\bar \alpha_{n,k}X^k\,, \qquad {\rm and}\qquad
d_n(X)=\sum_{n\ge 0} \frac{1}{k!}\bar d_{n,k}X^k\,,
\ea
with the convention $\bar \alpha_{0,0}=0$, and
\ba\label{eq:alpha0}
\bar \alpha_{0,1}=-\frac 12 \frac{m^2}{\Lambda^2}\,, \quad {\rm and} \quad
\bar\alpha_{2,0}=1+\frac 12 \bar \alpha_{1,0}\,.
\ea
If instead one had for instance a theory where $\alpha_0$ is constant and does not carry the mass term linear in $X$, \ie setting $\bar \alpha_{0,1}=0$ unlike what is indicated in \eqref{eq:alpha0}, would result in an infinitely strongly coupled vacuum solution  $\langle A_\mu \rangle =0$ which would be against the logic of the model presented here.

\subsection{Null Eigenvector}

There exist multiple ways to show that PN and GP are genuinely different theories (not even related by Vector dualities \cite{deRham:2014lqa}). It was first proven in \cite{deRham:2020yet} that their scattering amplitudes differ, hence establishing their inequivalent nature. It was then shown in \cite{deRham:2021efp} that the class of cosmological predictions of PN could differ from the  GP ones. However, the most immediate and natural hint at the fact that these theories are truly different is to envisage how the constraint is realized. This can be identified by considering the null eigenvector (NEV) of their respective Hessian matrices of field velocities. One of the underlying hypotheses of GP is that all modes in the decoupling limit should have at most second-order equations of motion, which is related to the fact that the component $A_0$ of the massive vector field remains non-dynamical when adding the GP interactions to the standard Proca theory. This corresponds to a NEV in the field-space direction $(1,\vec{0}\,)$. On the other hand, the very construction of the pure PN interactions inhibits such a clean identification of the non-propagating degree of freedom. Yet it was proven in \cite{deRham:2020yet} that their Hessian was still degenerate as it enjoys a null eigenvector. Since we shall be interested in the presence of a secondary constraint, it is beneficial to first identify the degenerate direction of (E)PN and hence review how the NEV can be identified.

To this end we first introduce the tensor
\begin{equation}
	\mathcal{Z}=\X^{-1} \eta^{-1} \,,
\end{equation}
which can be shown to be symmetric, $\Z=\Z^T$, and to satisfy the properties
\begin{align}
	\mathcal{Z}^{\alpha\beta} f_{\beta \gamma} &= \X\ud{\alpha}{\gamma} \,, \\
	\mathcal{Z}\mnup f_{\nu \alpha} \mathcal{Z}^{\alpha \beta} &= \eta^{\mu \beta} \,.
\end{align}
We further define a tensor $W^\mu{}_\nu$ via
\begin{equation}
	W\ud{\mu}{\nu} = \mathcal{Z}^{\mu \alpha} \p_{\alpha} \phi_{\nu} \,,
	\label{eq:defWtensor}
\end{equation}
which can be seen to belong to the Lorentz group,
\begin{equation}
	W^{\mu}{}_{\alpha}\eta^{\alpha\beta} W^{\nu}{}_{\beta} = \eta\mnup \,.
\end{equation}
The generalized Hessian matrix is given by
\begin{equation}
	\mathcal{H}^{\mu \nu, \alpha \beta} = \frac{\p^2 \L_{\rm EPN}}{\p \p_\mu {A}_{\alpha} \p \p_\nu {A}_{\beta}} = \mathcal{H}^{\nu \mu, \beta \alpha} \,,
\end{equation}
and $\mathcal{H}^{\mu \nu}\equiv \mathcal{H}^{00,\mu \nu}$ plays the role of the kernel of the kinetic term of the Lagrangian, hence encoding information about the propagating degrees of freedom of the theory. One can verify that the vector
\begin{equation}
	V_{\mu} = W\ud{0}{\mu} \,,
	 \label{eq:PN NEV constraints}
\end{equation}
is the normalized time-like NEV of the Hessian matrix of PN \cite{deRham:2020yet},
\begin{equation}
	\mathcal{H}\mnup V_{\mu} = 0 \,, \qquad V^{\mu} V_{\mu} = - 1 \,.
\end{equation}
In addition, as proven in \cite{deRham:2021efp}, the inclusion of the GP operators $\L_n[\p A]$ leaves the Hessian matrix $\mathcal{H}^{\mu \nu, \alpha \beta}$ invariant and can therefore be safely added to the PN Lagrangian without affecting the constraint structure, resulting in the EPN model.

\subsection{Pair of second class constraints}

The existence of a NEV in (E)PN theory implies the existence of a constraint, which must be second class as the theory does not have any gauge symmetries. The constraint, therefore, removes one phase-space variable corresponding to half a Lagrangian degree of freedom. The removal of the other half then necessitates the existence of another second class constraint. While the analyses of \cite{deRham:2020yet,deRham:2021efp} did not derive the latter, it was however argued that the Hessian constraint was enough to prove the absence of  the full Ostrogradsky ghost in (E)PN. This follows for multiple physically motivated reasons:
\begin{enumerate}
\item  First of all, since PN is a DL of generalized massive gravity (see \cite{deRham:2014gla,deRham:2016plk}) for which the secondary constraint was already derived in the literature (proven fully non-linearly in \cite{Hassan:2011hr,Golovnev:2011aa,Alexandrov:2013rxa,Golovnev:2017iix}), it directly follows that the secondary constraint has to be realized in PN (see \cite{deRham:2014zqa} on what it means physically to take a DL).
\item About the trivial vacuum $\langle A_\mu \rangle =0$, we recover a standard Proca theory at the linear level, which as is well known propagates $D-1$ degrees of freedom, implying that PN has to propagate {\it at least} $D-1$ degrees of freedom. Since it is expressed in terms of $D$ vector field components with only first derivatives acting on them at the level of the action, it can {\it at most} propagate $D$ modes. However, the presence of a NEV for the Hessian matrix implies that the system is degenerate and must in fact propagate strictly fewer than $D$ field space degrees of freedom. Moreover, as explained in \cite{deRham:2014zqa} and reviewed already in Section~\ref{sec:HalfDegree}, a local, Lorentz and parity preserving field theory cannot propagate an odd number of physical field space degrees of freedom. Together, these arguments imply that about any solution analytically connected to the standard Lorentz invariant vacuum $\langle A_\mu \rangle =0$, there should be precisely $D-1$ field space degrees of freedom. This result is consistent with previous arguments reminiscent of massive gravity \cite{Hassan:2011hr,Alexandrov:2013rxa}.
\end{enumerate}
The above points are also consistent with further analyses:
\begin{enumerate}\setcounter{enumi}{2}
\item Perturbations of (E)PN on cosmological backgrounds were analysed in \cite{deRham:2021efp}, where they were shown to exhibit the expected number of physical degrees of freedom, all of which were identified as being stable (even though following the logic of the analysis presented in \cite{ErrastiDiez:2022qvd} a mismatch would already have been identified at that level).
\item Positivity bounds, which rely on unitarity (which would be broken if half a ghost degree of freedom was propagating) together with Lorentz invariance and crossing symmetry were shown to be satisfied for generic (E)PN parameters in \cite{deRham:2022sdl}.
\end{enumerate}
These points suggest (if not prove) that the Hessian constraint in (E)PN is actually part of a pair of second class constraints fully removing the unwanted ghostly degree of freedom. Nevertheless, it remains an interesting open question to establish how the constraint structure manifests itself more precisely. In the rest of this work, we turn our attention to this question.

\subsection{Extended Proca-Nuevo in two dimensions}
\label{sec:EPN2dConstraint}

In this Section, we will focus on the example analysed in \cite{ErrastiDiez:2022qvd} corresponding to the EPN model in $D=2$ dimensions. This model is interesting because the structure of the Lagrangian can be made very explicit and allows for a tractable analytical treatment.

The most general two-dimensional EPN Lagrangian is given by \cite{deRham:2020yet}
\begin{equation}
\label{eq:LEPN2d}
	\L^{\rm (2d)}_{\rm EPN} = \Lambda^2 \( \alpha_0(X) +\alpha_1(X)  [\K] + \alpha_2(X) \left([\K]^2-[\K^2] \right) + \frac{d_1(X)}{\Lambda}[\p A] \) \,,
\end{equation}
where the analytic functions $\alpha_{0,1,2}(X)$ and $d_{1}(X)$ satisfy the expansion given in \eqref{eq:expansion_alpha_d} with the convention \eqref{eq:alpha0}. Note however that even though $\L_2[\p A]$ is multiplied by an arbitrary function of $A^2$, namely $d_2$, the whole term $d_2(X) ([\p A]^2-[(\p A)^2])$ is a total derivative and hence can be discarded \cite{BeltranJimenez:2019wrd}.

As discussed in \cite{deRham:2020yet}, in $D=2$ dimensions, the Lagrangian can be expressed in closed form in terms of the  following variables:
\begin{equation}
	x_{\pm} = 1 \pm 1 + \frac{A_1' \mp \dot{A}_0}{\Lambda} \,, \qquad y_{\pm} = \frac{\dot{A}_1 \mp A_0'}{\Lambda} \,, \qquad N_{\pm} = \sqrt{x_{\pm}^2 - y_{\pm}^2} \,,
\end{equation}
where a dot (prime) stands for time (space) derivative. We also define their reduced versions
\begin{equation}
	\bar{x}_{\pm} = \frac{x_{\pm}}{N_{\pm}} \,, \qquad \bar{y}_{\pm} = \frac{y_{\pm}}{N_{\pm}} \,,
\end{equation}
allowing us to rewrite the Lagrangian in the form
\begin{equation}
\label{eq:LEPN_2d}
	\L^{\rm (2d)}_{\rm EPN} = \Lambda^2 \left[ \tilde{\alpha}_0 + \tilde\alpha_1  N_{+}+ d_1  x_{+} + \frac{\alpha_2}{2} \(N_{+}^2 - N_{-}^2\) \right] \,,
\end{equation}
with $\tilde{\alpha}_0 = \alpha_0-2(\alpha_1+d_1)+2\alpha_2$ and $\tilde \alpha_1 = \alpha_1 - 2\alpha_2$.

In principle, there is another branch of solutions to the defining matrix square root equation, although as already discussed below Eq.~\eqref{eq:XX=f} we only commit to theories which are continuously related to the standard Proca one at linear order and so implicit in the expression for $\X\mupn=(\sqrt{\eta^{-1}f})\mupn$ is the choice of solution satisfying $\X\mupn[0]=\delta\mupn$. On the other hand, the theory corresponding to the other branch does not reduce to the standard Proca one on the trivial vacuum, and in fact, is not even properly formulated about the $\langle A_\mu\rangle =0$ vacuum and we do not consider it any further. Explicitly, the physical branch has the following perturbative expansion:
\begin{equation}
\begin{aligned}
	\L^{\rm (2d)}_{\rm EPN} =&- \frac14 F^2- \frac12 m^2 A^2 +\Lambda \p A\left( \left( \bar \alpha_{1,1}+\bar d_{1,1} \right)A^2+\frac{1}{8\Lambda^2} F^2 \right) \\
	&+\frac {\Lambda^2}2 \bar \alpha_{0,2}A^4+\frac 18 \left( \bar \alpha_{1,1} -2\bar \alpha_{2,1} \right)F^2 A^2+\frac{1}{128\Lambda^2}(F^2)^2-\frac{1}{16 \Lambda^2}F^2 (\p A)^2+\Lambda^2\mathcal{O}\(A^6,  \frac{\p A^5}{\Lambda}\)\,,
\end{aligned}
\end{equation}
where we use the notation $F^2 = F\mnup F\mn$ and $A^2=A^{\mu}A_{\mu}$.

The Hessian matrix is given fully non-linearly by
\begin{equation}
\label{eq:EPN2dHessian}
	\mathcal{H}\mnup \equiv \frac{\p \L^{\rm (2d)}}{\p \dot{A}_{\mu} \p \dot{A}_{\nu}} = - \frac{\tilde\alpha_1}{N_+} \begin{pmatrix}
	\bar{y}_+^2 &\bar{x}_{+} \bar{y}_{+} \\
	\bar{x}_{+} \bar{y}_{+} & \bar{x}_{+}^2
	\end{pmatrix} \,,
\end{equation}
whose determinant vanishes and thus admits a NEV. The latter can easily be inferred by inspection but can also be derived from the tensor $W\ud{\mu}{\nu}$, which here takes the form
\begin{equation}
	{W}\ud{\mu}{\nu} = \begin{pmatrix}
		 \bar{x}_{+} & - \bar{y}_{+} \\
		- \bar{y}_{+} & \bar{x}_{+}
	\end{pmatrix} \,.
\end{equation}
The NEV is then
\begin{equation}
	V_{\mu} =W^0{}_\mu = \left(\bar{x}_{+} , - \bar{y}_{+} \right) \,, \qquad \text{so that} \qquad V_{\mu} \eta\mnup V_{\nu} = - 1 \,,
\label{eq:VPN2d}
\end{equation}
and it is easy to check that it indeed annihilates the Hessian matrix,
\begin{equation}
	\mathcal{H}\mnup V_{\mu} = 0 \,.
\end{equation}
For later use we also introduce the vector $V^\perp$ normal to the NEV,
\begin{equation}
\label{eq:EPN2dNEV}
V^\perp_\mu =  W^1{}_\mu= \left( - \bar{y}_{+} , \bar{x}_{+} \right) \,, \qquad \text{so that} \qquad
V^\perp_\mu \eta\mnup V^\perp_\nu = 1 \qquad \text{and} \qquad V_{\mu} \eta\mnup V^\perp_\nu = 0 \,.
\end{equation}
The Euler-Lagrange equations for EPN in $D=2$ have the form
\begin{equation}
	\mathcal{E}^{\mu} \equiv \mathcal{H}\mnup \ddot{A}_{\nu} + u^{\mu} = 0 \,,
	\label{eq:EL}
\end{equation}
where the explicit expression for the acceleration-free part $u^{\mu}$ is provided in Appendix \ref{app:EPNLagrangian}. One can now contract the equation $\mathcal{E}^{\mu}$ with the vector $V_{\mu}$ and make use of the fact that the latter is the NEV of the Hessian matrix to find a constraint
\begin{equation}
	\mathcal{C}_{V1} \equiv V_{\mu} \mathcal{E}^{\mu} = V_{\mu} u^{\mu} \approx 0 \,.
	\label{eq:secondclassEVa}
\end{equation}
Here and in what follows, the symbol ``$\approx$" is used to designate ``on the constraint surface''.  It is also possible to show that the constraint takes the following compact form
\begin{equation}
	\mathcal{C}^{\rm (EPN)}_{V1} = \Lambda^2 \left[ \Phi_1 \tilde\alpha_1 + \phi_0 ( \tilde\alpha_{0,X} + d_{1,X} ) + \phi_1 \tilde\alpha_{1,X} + \phi_2 \left( \alpha_{2,X} + \frac12 d_{1,X} \right) \right] \approx 0 \,,
	\label{eq:secondclassEVb}
\end{equation}
and we refer the reader to Appendix \ref{app:EPNLagrangian} for the definitions of the functions $\phi_i$ and $\Phi_1$.\\

This establishes the existence of a primary constraint, as was already proven to be true in two and four dimensions in \cite{deRham:2020yet, deRham:2021efp}. The extension of the proof to arbitrary dimensions is straightforward. The following sections are devoted to the proof that the model also has a secondary constraint, leaving precisely one dynamical degree of freedom in $D=2$ dimensions (or $D-1$ degrees of freedom in $D$ dimensions), thus avoiding the existence of any half degree of freedom in this local, Lorentz and parity invariant theory.

\section{Constraint analysis in the Lagrangian picture}
\label{sec:2dConstraint}
\label{sec:LagrConstraint}

In this Section, we demonstrate the existence of a secondary second class constraint using the Lagrangian formalism, before turning to the formal computation of the constraint algebra of the system using the Hamilton-Dirac formalism in Section \ref{sec:HamConstraint}. In each case, we begin with the analysis of linear Proca theory for the sake of pedagogy and to fix notation, followed by a warm-up treatment of GP theory, before turning our attention onto the EPN model of interest.

\subsection{Linear Proca}
\label{ssec:ProcaLagrConstraint}

The linear Proca theory in two dimensions has the Lagrangian,
\ba
	\L^{\rm (2d)}_{\rm Proca} = - \frac{1}{4} F^2 - \frac12 m^2 A^2 = \frac12 (\dot{A}_1-A_0')^2 - \frac12 m^2 (-A_0^2 + A_1^2) \,.
\ea
The Lagrangian is independent of $\dot{A}_0$, hence $A_0$ is non-propagating and we immediately obtain the following expressions for the Hessian matrix and NEV:
\begin{equation}
	\mathcal{H}\mnup = \begin{pmatrix}
	0 & 0 \\
	0 & 1
	\end{pmatrix} \,, \qquad V_{\mu} = (1,0) \,, \qquad V^\perp_{\mu}=(0,1) \,.
\end{equation}
The Euler-Lagrange equations read
\begin{equation}
\label{eq:EmuProca}
	\mathcal{E}^{\mu} = \mathcal{H}\mnup \ddot{A}_{\nu} + u^{\mu} = 0 \,,
\end{equation}
with
\begin{equation}
	u^{\mu} = \begin{pmatrix}
	A_0'' - m^2 A_0 - \dot{A}_1' \\
	m^2 A_1 - \dot{A}_0'
	\end{pmatrix} \,,
\end{equation}
and the primary constraint is given by
\begin{equation}
	\mathcal{C}_{V1} \equiv V_{\mu} u^{\mu} = A_0'' - m^2 A_0 - \dot{A}_1' \,.
\end{equation}
Consistency of the primary constraint under time evolution yields a secondary constraint\footnote{The constraints presented here agree with the well-known textbook results; see for instance \cite{Henneaux:1992ig} for electromagnetism (the massive case is given as an exercise) and \cite{Banerjee:1994pp}. These however differ from those given in \cite{ErrastiDiez:2020dux}.}
\begin{equation}
	\dot{\mathcal{C}}_{V1} = - \left( V^{\perp}_{\mu} \mathcal{E}^{\mu} \right)' + m^2 (A_1' - \dot{A}_0) \approx 0 \,,
\end{equation}
which, on-shell and for $m^2\neq0$, gives
\begin{equation}
	\mathcal{C}_{V2} \equiv A_1' - \dot{A}_0 \approx 0 \,,
\end{equation}
the familiar Lorenz condition which is now not a gauge choice but a constraint. We see that, once initial conditions are specified for $A_1$ and its time derivative, the variable $A_0$ is then completely determined (modulo spatial boundary conditions) and is therefore non-dynamical. This leaves us with a single Lagrangian degree of freedom in two dimensions.\\

We can now only make use of integrations by part to show that our Lagrangian can be recast in a way that explicitly shows that the associated phase space will simply be $\{ A_1 , p^1 \}$ where $p^1$ is the conjugate momentum to $A_1$.
\begin{align}
	\L^{\rm (2d)}_{\rm Proca} 	=& \frac12 ( A_0' - \dot{A}_1 )^2 + \frac12 m^2 (A_0^2 - A_1^2) \nonumber \\
	=& \frac12 \dot{A}_1^2 - \frac12  ( A_1')^2 - \frac12 m^2 A_1^2 - \frac12 A_0 \mathcal{C}_{V1}^{\rm (Proca)} + \frac12 A_1' \mathcal{C}_{V2}^{\rm (Proca)} + (\text{total derivatives}) \,.
\end{align}
Note that the Lagrangian includes two linear Lagrange multipliers $A_0$ and $A_1'$ and hence simply reduces to $\frac12 \dot{A}_1^2 - \frac12  ( A_1')^2 - \frac12 m^2 A_1^2$ on the constraint surface. It is now independent of both $A_0$ and $\dot{A}_0$ and hence the dynamics will be fully fixed by only specifying $2$ initial conditions for $A_1$ and $\dot{A}_1$ (or its conjugate momentum $p^1$ equivalently in the Hamiltonian formalism). This shows that there is only one propagating physical degree of freedom in the standard Proca model in $D=2$ dimensions.

\subsection{Generalized Proca}
\label{ssec:GenProcaLagrConstraint}

Next, we review the constraint analysis of GP \cite{Heisenberg:2014rta}. In $D=2$ the Lagrangian reads
\ba
	\L^{\rm (2d)}_{\rm GP}  = - \frac{1}{4} F^2 - \frac12 m^2 A^2 + \Lambda^2 d_0(X) + \Lambda d_1(X) \p_\alpha A^\alpha  \,.
\ea
For a generic functions $d_{1}$, the Lagrangian is no longer independent of $\dot{A}_0$, yet it is still true that the equations of motion are independent of $\ddot{A}_0$, hence $A_0$ is non-propagating. The resulting Hessian matrix, associated NEV and normal vector take again the exact same form as for the free Proca theory
\begin{equation}
	\mathcal{H}\mnup = \begin{pmatrix}
	0 & 0 \\
	0 & 1
	\end{pmatrix} \,, \qquad V_{\mu} = (1,0) \,, \qquad V^{\perp}_{\mu} = (0,1) \,.
\end{equation}
The Euler-Lagrange equations are given as in \eqref{eq:EmuProca}
\begin{equation}
\label{eq:EmuGP}
	\mathcal{E}^{\mu} = \mathcal{H}\mnup \ddot{A}_{\nu} + u^{\mu} = 0 \,,
\end{equation}
with now
\begin{equation}
	u^{\mu} = \begin{pmatrix}
	A_0'' - m^2 A_0 -\dot{A}_1' + 2 \Lambda^2 d_{0,X} A_0 + 2 \Lambda d_{1,X} \left( A_0 A_1' - A_1 \dot{A}_1 \right) \\
	m^2 A_1 - \dot{A}_0' - 2 \Lambda^2 d_{0,X} A_1 - 2 \Lambda d_{1,X} \left( A_0 A_0' - A_1 \dot{A}_0 \right)
	\end{pmatrix} \,,
\end{equation}
and the constraint spells
\begin{equation}
	\mathcal{C}_{V1} \equiv V_{\mu} u^{\mu} = A_0'' - m^2 A_0 - \dot{A}_1' + 2 \Lambda^2 d_{0,X} A_0 + 2 \Lambda d_{1,X} \left( A_0 A_1' - A_1 \dot{A}_1 \right) \approx 0 \,.
\end{equation}
Taking the time derivative of this constraint, it is straightforward to show that all second time derivatives of the fields can be eliminated  using combinations of the equations of motion, so that
\begin{equation}
	\dot{\mathcal{C}}_{V1} + V^{\perp}_{\mu} (\mathcal{E}^{\mu})' + 2 \Lambda A_1 d_{1,X} V^{\perp}_{\mu} \mathcal{E}^{\mu} \equiv \mathcal{C}_{V2} \approx 0 \,,
\end{equation}
where $\mathcal{C}_{V2}$ is free of any higher-order time derivatives (or accelerations) and does \textit{not} vanish on the primary constraint surface. Hence it is a genuinely new second class constraint. Its explicit expression reads
\begin{equation}
\begin{aligned}
	\mathcal{C}_{V2} =&\; (m^2 - 2\Lambda^2 d_{0,X}) (\p_{\mu} A^{\mu} + 2 \Lambda d_{1,X} A_1^2) - 4 \Lambda^2 A_1 d_{1,X}^2 (A_0 A_0' - A_1 \dot{A}_0)  \\
	&- 2 \Lambda d_{1,X} \left[ (\p_{\mu} A^{\mu})^2 - \p_{\mu} A_{\nu}\p^{\mu} A^{\nu} + A_0 \left( A_0' - \dot{A}_1 \right)' \right] - 4\Lambda^2 d_{0,XX} A_{\mu} A_{\nu} \p^{\mu} A^{\nu} \\
	&- 4 \Lambda d_{1,XX} A_{\mu} A_{\nu} \left[ \p^{\mu} A^{\nu} \p_{\sigma} A^{\sigma} - \p_{\sigma} A^{\mu} \p^{\sigma} A^{\nu} \right] \,.
\end{aligned}
\end{equation}
This is sufficient to establish the consistency of GP theory from the point of view of the constraint structure.

\subsection{Extended Proca-Nuevo}
\label{ssec:EPNLagrConstraint}
We finally turn to the analysis of EPN in $D=2$ dimensions. The primary constraint, Hessian matrix, NEV and normal vector are given above in Section \ref{sec:EPN2dConstraint}, see Eqns.~(\ref{eq:EPN2dHessian}--\ref{eq:EPN2dNEV}). Following the previous warm-up examples, the strategy is now clear: Take the time derivative of the primary constraint and add combinations of the equations of motion so as to remove all field accelerations (and derivatives thereof). The result is the secondary constraint.

\paragraph{Secondary Constraint for General EPN:}
Carrying out this procedure we obtain
\begin{equation}
\label{eq:C22dEPN}
	\dot{\mathcal{C}}_{V1} + V^{\perp}_{\mu}(\mathcal{E}^{\mu})' + 2 \frac{\Lambda}{\tilde\alpha_1} \beta \left( V^{\perp}_{\mu} \mathcal{E}^{\mu} \right) \equiv \mathcal{C}_{V2} \approx 0 \,,
\end{equation}
where
\begin{equation}
\begin{aligned}
	\beta =&\; (\bar{y}_+ A_0 + \bar{x}_+ A_1) \left( \tilde\alpha_{0,X} + \tilde{d}_{1,X} \right) -(\Delta A_0 - (1-\Sigma)A_1) \tilde\alpha_{1,X} \\
	& - \left[ (\bar{x}_+ \Delta - \bar{y}_+ \Sigma) A_0 + (\bar{x}_+ \Sigma - \bar{y}_+ \Delta) A_1 \right] \left( 2 \tilde\alpha_{2,X} + \tilde{d}_{1,X} \right) \,.
\end{aligned}
\end{equation}
The exact expression for the secondary constraint $\mathcal{C}_{V2}$  is given in the minimal model below. For the generic EPN, while straightforward to derive, its exact expression is rather formidable and not particularly illuminating, we thus refer the reader to Appendix~\ref{app:EPNSecConstraint} for its full expression. We emphasize however that it is a true independent constraint: it does not involve any accelerations and does not vanish on the primary constraint surface.

We conclude that EPN theory possesses a pair of constraints that together remove a full Lagrangian degree of freedom, since there are no gauge symmetries, thus defining a consistent massive spin-1 model. While this proof applies only in $D=2$ dimensions, a partial proof in generic dimension will be given in Section \ref{sec:HigherDConstraint}.

\paragraph{Minimal Model:} As a special example, we can consider the minimal model which will also be studied in generic dimensions in Section~\ref{sec:HigherDConstraint}. Focusing for now in $D=2$ dimensions, the minimal model corresponds to setting $d_1(X)=\alpha_2(X)\equiv0$ and $\alpha_1(X)\equiv 2$  in the EPN Lagrangian \eqref{eq:LEPN2d}, while keeping the potential arbitrary $\alpha_0(X)=-\frac 12 m^2 X + V(X)$. Note that it would not make sense to set $\alpha_0$ to a constant as it would set $m=0$ and the field would lose its mass term on the trivial vacuum, leading to an infinitely strong coupling. The minimal model in two dimensions is then given by
\ba
\L^{\rm (2d)}_{\rm Minimal}=\Lambda^2\(\alpha_0(X)-2[\mathcal{K}]\)=\Lambda^2\(\alpha_0(X)-2[\mathcal{X}]+4\)\,.
\ea
As in the previous cases, the Euler-Lagrange equations are given by $\mathcal{E}^{\mu} = \mathcal{H}\mnup \ddot{A}_{\nu} + u^{\mu} = 0 $ with the vector $u^\mu$ now given by (see Appendix~\ref{app:EPNLagrangian})
\ba
u^\mu_{\rm Minimal}=-\frac{2}{N_+^3} \begin{pmatrix}
 x_{+} y_{+} \left( 2  \dot{A}_0' -  A_1'' \right) - x_{+}^2  A_0'' + \left( x_{+}^2 + y_{+}^2 \right)  \dot{A}_1' \\
 x_{+} y_{+} \left( 2  \dot{A}_1' -  A_0'' \right) - y_{+}^2  A_1'' + \left( x_{+}^2 + y_{+}^2 \right)  \dot{A}_0'
\end{pmatrix}-2\Lambda^2 \alpha_{0,X}A^\mu\,.
\ea
In this case, the primary constraint is given by
\begin{equation}
	\mathcal{C}^{\rm (Minimal)}_{V1} = -2 \Lambda \left( \bar{x}_{+} \p_1 \bar{y}_{+} - \bar{y}_{+} \p_1 \bar{x}_{+}  \right) + 2 \alpha_{0,X} \Lambda^2 \left( \bar{x}_{+} A_0 + \bar{y}_{+} A_1 \right)   \approx 0 \,,
	\label{eq:minimalCV1}
\end{equation}
and  the secondary constraint $\mathcal{C}_{V2}$  takes the form (see \eqref{eq:C22dEPN}),
\begin{equation}
\begin{aligned}
\Lambda^{-3}\mathcal{C}^{\rm (Minimal)}_{V2}=&- \frac{2}{\Lambda^2} \left( \bar{x}_{+} \p_1 \bar{y}_{+} - \bar{y}_{+} \p_1 \bar{x}_{+}  \right)^2  \\
&+ 2 \alpha_{0,X} \left(2\bar{x}_+-N_+\right) +2 \alpha_{0,X}^2 (\bar{y}_+ A_0 + \bar{x}_+ A_1) ^2 \\
&+ \frac{2}{\Lambda}\alpha_{0,XX} \left(\dot{X}\left( \bar{x}_{+} A_0 + \bar{y}_{+} A_1 \right) - 2\(\bar{y}_+ A_0 + \bar{x}_+ A_1\)  \left(-A_0 A_0'+A_1 A_1'\right)\right) \,.
\end{aligned}
\end{equation}
Even in the case where  the potential reduces to a quadratic mass term, $\alpha_0=-\frac 12 m^2 A_\mu^2$ and $\alpha_{0,XX}=0$, we see that both constraints are independent. Once again,  the minimal massless case $\alpha_{0,X}\equiv 0$ is infinitely strongly coupled and meaningless.

\section{Constraint analysis in the Hamiltonian picture}
\label{sec:HamConstraint}

In this Section, we carry out the Hamilton-Dirac analysis of the two-dimensional EPN theory. We will demonstrate that the model enjoys a pair of second class constraints, leaving a two-dimensional reduced phase space, or equivalently a single Lagrangian degree of freedom. We warm up again with the examples of linear Proca and GP.

\subsection{Linear Proca}
\label{ssec:ProcaHamConstraint}

We perform a 1+1 decomposition of the linear Proca Lagrangian density,
\beq
\L^{\rm (2d)}_{\rm Proca}=\frac{1}{2}\left(\dot{A}_1-A_0'\right)^2+\frac{m^2}{2}\left(A_0^2-A_1^2\right) \,,
\eeq
so the canonical momenta read
\beq
p^0=0 \,,\qquad p^1=\dot{A}_1-A_0' \,,
\eeq
and we infer the primary constraint
\beq
\mathcal{C}_1\equiv p^0\approx 0 \,.
\eeq

The canonical or base Hamiltonian density reads
\begin{equation}
\begin{aligned}
	\mathcal{H}_{\rm base}&= p^{\mu}\dot{A}_{\mu}-\L \\
&=\frac{1}{2}(p^1)^2+p^1A_0'-\frac{m^2}{2}\left(A_0^2-A_1^2\right) \,,
\end{aligned}
\end{equation}
and the ``augmented'' or ``primary'' Hamiltonian is obtained by adding the primary constraints with arbitrary Lagrange multipliers. In this case, $\mathcal{H}_{\rm aug}=\mathcal{H}_{\rm base}+\lambda_1\mathcal{C}_1$.

Consistency of the constraint $\mathcal{C}_1$ under time evolution, $\dot{\mathcal{C}}_1\approx0$,  may either fix the Lagrange multiplier $\lambda_1$ or else produce a secondary constraint\footnote{
There is a possibility that the secondary constraint is simply inconsistent with the primary one, signaling a fundamentally pathological theory. To our knowledge, examples of this kind are all contrived or trivial and the theory can be seen to be inconsistent without performing a constraint analysis. In any case, this outcome will not occur for the models we investigate here.}.
The latter happens when the Poisson bracket $\{\mathcal{C}_1(x),\mathcal{C}_1(y)\}$ vanishes weakly, as is obviously the case for linear Proca\footnote{It is worth emphasizing that the vanishing of the Poisson bracket of a constraint with itself is not automatic. An example of a ``self-second class'' constraint arises in Lorentz-breaking Ho\v{r}ava-Lifshitz gravity \cite{Li:2009bg}.}. Thus we obtain a secondary constraint:
\beq\bal
\dot{\mathcal{C}}_1&= \{\mathcal{C}_1,H_{\rm aug}\} \\
&=(p^1)'+m^2A_0 \qquad \Rightarrow\qquad \mathcal{C}_2\equiv (p^1)'+m^2A_0\approx0\,.
\eal\eeq
We observe that $\mathcal{C}_1$ and $\mathcal{C}_2$ are second class as they do not commute with each other,
\beq
\{\mathcal{C}_1(x),\mathcal{C}_2(y)\}=-m^2\delta(x-y) \,.
\eeq
Preservation in time of $\mathcal{C}_2\approx0$ thus fixes the multiplier $\lambda_1$,
\beq
\dot{\mathcal{C}}_2=-m^2A_1'+m^2\lambda_1\approx0 \qquad \Rightarrow\qquad \lambda_1=A_1' \,.
\eeq
Finally, the total Hamiltonian from which the equations of motion are derived (which must be supplemented with initial conditions consistent with \textit{all} the constraints) is obtained by substituting the solutions for the Lagrange multipliers into the augmented Hamiltonian density,
\beq
\mathcal{H}_{\rm tot}=\mathcal{H}_{\rm base}+p^0A_1' \,.
\eeq
Since each second class constraint reduces the dimensionality of the physical phase by unity, we are left in the end with a two-dimensional phase space or a single degree of freedom in field space.

\subsection{Generalized Proca}
\label{ssec:GenProcaHamConstraint}

We refer the interested reader to \cite{Heisenberg:2014rta,BeltranJimenez:2019wrd} for a more complete constraint analysis of GP, while here we content ourselves with the derivation of the primary and secondary constraints, in addition to the proof that they are second class. The 1+1-decomposed GP Lagrangian density is
\beq
\L^{\rm (2d)}_{\rm GP}=\frac{1}{2}\left(\dot{A}_1-A_0'\right)^2+\frac{m^2}{2}\left(A_0^2-A_1^2\right)+\Lambda^2d_0(X)+\Lambda d_1(X)\left(-\dot{A}_0+A_1'\right) \,.
\eeq
From the canonical momenta,
\beq
p^0=-\Lambda d_1(X) \,,\qquad p^1=\dot{A}_1-A_0' \,,
\eeq
we infer the primary constraint $\mathcal{C}_1\equiv p^0+\Lambda d_1(X)\approx 0$ as well as the base and augmented Hamiltonian densities,
\beq\bal
\mathcal{H}_{\rm base}&= \frac{1}{2}(p^1)^2+p^1A_0'-\frac{m^2}{2}\left(A_0^2-A_1^2\right)-\Lambda^2d_0(X)-\Lambda d_1(X)A_1' \,,\\
\mathcal{H}_{\rm aug}&= \mathcal{H}_{\rm base}+\lambda_1\mathcal{C}_1 \,.
\eal\eeq
Although less obvious in this case, it can again be easily checked that $\mathcal{C}_1$ commutes with itself, so its consistency under time evolution generates a secondary constraint:
\beq\bal
\mathcal{C}_2\equiv \dot{\mathcal{C}}_1 &= \{\mathcal{C}_1,H_{\rm aug}\} \\
&=(p^1)'+m^2A_0-2\Lambda^2d_{0,X}A_0+2\Lambda d_{1,X}\left(-A_0A_1'+A_1A_0'+p^1A_1\right)\approx0 \,.
\eal\eeq

It is clear that $\mathcal{C}_1$ and $\mathcal{C}_2$ do not Poisson-commute (since they do not in the case of linear Proca, and the GP functions $d_{0,1}$ are generic), although we will not derive the explicit result. It follows that they are independent, second class constraints, implying the absence of further constraints and the determination of the Lagrange multiplier $\lambda_1$ in terms of the phase space variables, and hence of the total Hamiltonian.

\subsection{Extended Proca-Nuevo}
\label{ssec:EPNHamConstraint}

The 1+1-decomposed EPN Lagrangian was already given above in \eqref{eq:LEPN_2d}. Working out the canonical momenta, we find
\beq\bal
p^0&= -\Lambda\left[\tilde{\alpha}_1\bar{x}_{+}+2\alpha_2\Sigma+d_1\right] \,,\\
p^1&= -\Lambda\left[\tilde{\alpha}_1\bar{y}_{+}+2\alpha_2\Delta\right] \,,
\eal\eeq
with $\Sigma=1+A_1'/\Lambda$ and $\Delta=-A_0'/\Lambda$.
We infer the following primary constraint:
\beq
\mathcal{C}_1\equiv \left[\frac{p^0}{\Lambda}+2\alpha_2\Sigma+d_1\right]^2-\left[\frac{p^1}{\Lambda}+2\alpha_2\Delta\right]^2-\tilde{\alpha}_1^2 \approx0 \,.
\eeq
This is an interesting novelty of (E)PN theory relative to GP: the primary constraint is non-linear in the momenta\footnote{Of course any function of a constraint is also a constraint, defining the same hypersurface in phase space. However, some caution is needed when dealing with a constraint which is non-linear in all variables. Namely, one must check the so-called regularity condition, i.e.\ the requirement that the Jacobian of the constraints must have constant rank throughout phase space \cite{Henneaux:1992ig}. It can be easily checked that the regularity condition is satisfied by the constraint $\mathcal{C}_1$.}.
The resulting base and augmented Hamiltonians are therefore
\beq\bal
\mathcal{H}_{\rm base}&= \Lambda^2\left[\frac{p^0}{\Lambda}(1+\Sigma)-\frac{p^1}{\Lambda}\Delta-\tilde{\alpha}_0+2\alpha_2(\Sigma^2-\Delta^2)\right] \,,\\
\mathcal{H}_{\rm aug}&= \mathcal{H}_{\rm base}+\lambda_1\mathcal{C}_1 \,.
\eal\eeq
The next question is whether $\mathcal{C}_1$ commutes with itself. Since the consistency of the theory hinges on this question, we shall provide explicit details. Observe first that terms arising from derivatives of $\tilde{\alpha}_{1}$, $\alpha_2$ and $d_1$ yield zero. Indeed, such terms do not involve spatial derivatives of the field and hence give a contribution of the form
\beq
\{\mathcal{C}_1(x),\mathcal{C}_1(y)\}\supset \int \d z\left[F(x,y)\delta(x-z)\delta(y-z)-(x\leftrightarrow y)\right]=\left[F(x,y)-F(y,x)\right]\delta(x-y) \,,
\eeq
which vanishes, as can be seen more explicitly by integrating with an arbitrary test function. Contributions from the remaining terms give
\ba
\{\mathcal{C}_1(x),\mathcal{C}_1(y)\}
&=&\int \d z\bigg\{ 8\left[\alpha_2\left(\frac{p^1}{\Lambda}+2\alpha_2\Delta\right)\right]_x\left(\frac{p^0}{\Lambda}
+2\alpha_2\Sigma+d_1\right)_y\delta'(x-z)\delta(y-z) \nn \\
&&\phantom{\int \ \ }-8\left[\alpha_2\left(\frac{p^0}{\Lambda}+2\alpha_2\Sigma+d_1\right)\right]_x
\left(\frac{p^1}{\Lambda}+2\alpha_2\Delta\right)_y\delta'(x-z)\delta(y-z)\nn\\
&&\phantom{\int  \ \ }-(x\leftrightarrow y) \bigg\} \nn \\
&=&8\delta'(x-y)\bigg\{ \left[\alpha_2\left(\frac{p^1}{\Lambda}+2\alpha_2\Delta\right)\right]_x\left(\frac{p^0}{\Lambda}+2\alpha_2\Sigma+d_1\right)_y \nn \\
&&\quad -\left[\alpha_2\left(\frac{p^0}{\Lambda}+2\alpha_2\Sigma+d_1\right)\right]_x\left(\frac{p^1}{\Lambda}+2\alpha_2\Delta\right)_y \bigg\} -(x\leftrightarrow y) \nn \\
&=&8\delta'(x-y)\left[F(x,y)+F(y,x)\right] \,,
\ea
where now
\beq
F(x,y)\equiv \alpha_2(x)P^1(x)P^0(y)-\alpha_2(x)P^0(x)P^1(y) \,,
\eeq
\beq
P^0\equiv \frac{p^0}{\Lambda}+2\alpha_2\Sigma+d_1 \,,\qquad P^1\equiv \frac{p^1}{\Lambda}+2\alpha_2\Delta \,.
\eeq
Integrating with a test function, we get
\beq
\int \d yf(y)\{\mathcal{C}_1(x),\mathcal{C}_1(y)\}=8f(x)\left[F^{(1)}(x,x)+F^{(2)}(x,x)\right] \,,
\eeq
where $F^{(n)}$ denotes differentiation w.r.t.\ to the $n$-th argument of the function $F$. It is now easy to see that $F^{(2)}(x,x)=-F^{(1)}(x,x)$, establishing the consistency of the constraint $\mathcal{C}_1\approx 0$ under time evolution. Thus we obtain the secondary constraint
\begin{align}
\mathcal{C}_2 \equiv& \; \frac{\dot{\mathcal{C}}_1}{\Lambda}=\frac{1}{\Lambda}\{\mathcal{C}_1,H_{\rm aug}\}=\frac{1}{\Lambda}\{\mathcal{C}_1,H_{\rm base}\} \\
=& \; 2\left(\frac{p^0}{\Lambda}+2\alpha_2\Sigma+d_1\right) \left[\frac{\{p^0,H_{\rm base}\}}{\Lambda^2}-2\left(2\alpha_{2,X}\Sigma+d_{1,X}\right)\left(A_0(1+\Sigma)+A_1\Delta\right)-2\alpha_2\frac{\Delta'}{\Lambda}\right] \nn  \\
&- 2\left(\frac{p^1}{\Lambda}+2\alpha_2\Delta\right)\left[\frac{\{p^1,H_{\rm base}\}}{\Lambda^2}-4\alpha_{2,X}\Delta\left(A_0(1+\Sigma)+A_1\Delta\right)-2\alpha_2\frac{\Sigma'}{\Lambda}\right]  \\
&+ 4\tilde{\alpha}_1\tilde{\alpha}_{1,X}\left(A_0(1+\Sigma)+A_1\Delta\right) \vphantom{\frac{p^0}{\Lambda}} \,,\nn
\end{align}
and
\beq\bal
\{p^0,H_{\rm base}\}&= \Lambda^2\left[-2\tilde{\alpha}_{0,X}A_0+4 \alpha_{2,X}A_0\left(\Sigma^2-\Delta^2\right)\right]+\left(p^1+4\Lambda \alpha_2\Delta\right)' \,,\\
\{p^1,H_{\rm base}\}&=\Lambda^2\left[2\tilde{\alpha}_{0,X}A_1-4 \alpha_{2,X}A_1\left(\Sigma^2-\Delta^2\right)\right]+\left(p^0+4\Lambda \alpha_2\Sigma\right)' \,.
\eal\eeq
To complete the analysis of the constraint algebra it remains to verify the absence of tertiary constraint and the second class nature of $\mathcal{C}_1$ and $\mathcal{C}_2$. This is a straightforward\footnote{Let us remind the reader of the argument given in Section \ref{sec:reviewEPNConstraint} that the theory must propagate \textit{at least} $D-1$ degrees of freedom in view of the fact that (E)PN reduces to linear Proca theory upon linearization about the trivial vacuum. We therefore should not expect the presence of tertiary constraints, and given the absence of gauge symmetries, $\mathcal{C}_1$ and $\mathcal{C}_2$ must be second class and have non-zero Poisson bracket among each other.} but cumbersome task, so for brevity let us consider a minimal PN model with $\tilde{\alpha}_0,\alpha_2,d_1=0$ and $\tilde{\alpha}_1\propto X$. In this case, we find
\begin{align}
	\{\mathcal{C}_1(x),\mathcal{C}_2(y)\} =& \; \frac{8\tilde{\alpha}_{1,X}}{\Lambda}\left[2\tilde{\alpha}_{1,X}\left( A_0(1+\Sigma)+A_1\Delta\right)\left(p^0A_0+p^1A_1\right) -\tilde{\alpha}_1\left(p^0(1+\Sigma)+p^1\Delta\right)\right]\delta(x-y) \nn \\
&+ \frac{8\tilde{\alpha}_1\tilde{\alpha}_{1,X}}{\Lambda^2}\left[\left((p^0)'A_1+(p^1)'A_0\right)\delta(x-y)+2\left(p^0A_1+p^1A_0\right)\delta'(x-y)\right] \,,
\end{align}
which is not weakly zero. The final tally of degrees of freedom is now the familiar one: in two dimensions we have four phase space variables, reduced by two due to the presence of two second class constraints, hence a single Lagrangian degree of freedom.

\section{Minimal PN model in arbitrary dimensions}
\label{sec:HigherDConstraint}

In arbitrary spacetime dimension, it is unfortunately not possible to avoid the matrix square root structures that define (E)PN theory, making the problem of its constraint analysis technically more challenging.
The derivation of the primary and secondary constraint in massive gravity was derived in generality in arbitrary dimensions in \cite{Hassan:2011hr,Hassan:2011ea} and since PN follows the same structure as massive gravity, the same logic will apply. In what follows we restrict our attention to the minimal $D$-dimensional PN model given by the Lagrangian
\beq \label{eq:LEPND}
\L=\Lambda^D \left( \alpha_0(X) - 2 [\K] \right)=\Lambda^D \left( \alpha_0(X) - 2 [\X]+2D \right) \,,
\eeq
where we chose $\alpha_1 = -2$ to recover the canonical kinetic term $- \frac14 F\mnup F\mn$ at quadratic order in perturbation and the pure potential term $\alpha_0(X)$ should at the very least include the mass term, $\Lambda^D \alpha_0(X)=-\frac 12 m^2 A^2+\cdots$ and hence $\alpha_{0,X}\not \equiv 0$, (in fact if the theory admits a solution where $\alpha_{0,X}=0$ at some point in spacetime, then the theory is infinitely strongly coupled at that point and the classical solution cannot be trusted in the vicinity of that point). In the following we present the constraint analysis of this minimal model both in the Lagrangian and Hamiltonian pictures, finding in both cases the presence of a pair of constraints.

\subsection{Lagrangian picture}

Let us first collect some preliminary results. Given the definition of $W\ud{\mu}{\nu}$ in Eq.~\eqref{eq:defWtensor}, it is possible to show that
\begin{equation} \label{eq:trace derivative identity}
	\frac{\p \left[ \X^n \right]}{\p (\p_{\alpha} A_{\beta})} = \frac{n}{\Lambda^{D/2}} (\X^{n-1})^{\alpha}_{\phantom{\alpha}\mu} W^{\mu \beta} \,,
\end{equation}
which will be useful for $n=1$ here. From the definition of the generalized Hessian, applied to the Lagrangian \eqref{eq:LEPND}, we infer the relations
\begin{equation}
\label{eq:HW}
	\p_{\mu} W^{\alpha\beta} = -\frac{1}{2 \Lambda^{D/2}} \mathcal{H}^{\nu \alpha, \rho \beta} \p_{\mu} \p_{\nu} A_{\rho} \,,\qquad \mathcal{H}^{\hat{\mu}\nu, \alpha \beta} W\ud{\hat{\mu}}{\alpha} =\mathcal{H}^{\mu  \hat{\nu}, \alpha \beta} W\ud{\hat{\nu}}{\beta} =0 \,,
\end{equation}
where in the last expression the indices $\hat{\mu}$ and $\hat{\nu}$ have a fixed value and are not summed over. In particular this implies that the vector $V_\beta=W^0{}_\beta$ is a NEV for all $\mathcal{H}^{\mu 0, \alpha\beta}$ (\ie for any $\mu$, $\alpha$),
\ba
\label{eq:HW2}
\mathcal{H}^{\mu 0, \alpha\beta} V_\beta=0\,\quad\forall \ \mu, \alpha\,.
\ea

To proceed further, it will prove useful to write the equation of motion in two ways as
\beq
\mathcal{E}^{\mu} = \mathcal{H}\mnup \ddot{A}_{\nu} + u^{\mu} =  -2 \Lambda^{D/2} \dot{V}^{\mu} + \tilde{u}^{\mu} \,,
\eeq
where
\begin{align}
	u^{\mu} &= \p_j \dot{A}_{\alpha} \mathcal{H}^{0j,\mu \alpha} -2 \Lambda^{D/2} \p_i W^{i\mu} - 2\Lambda^2 \alpha_{0,X} A^{\mu} \,, \\
	\tilde{u}^{\mu} &= -2 \Lambda^{D/2} \p_i W^{i\mu} - 2\Lambda^2 \alpha_{0,X} A^{\mu} \,.
\end{align}
Given the NEV $V_{\mu}$, we infer the constraint
\begin{equation}
	\mathcal{C}_{V1} = V_{\mu}\mathcal{E}^{\mu} = V_{\mu} u^{\mu} = V_{\mu} \tilde{u}^{\mu} \,.
\end{equation}
The next step is to calculate the time derivative of $\mathcal{C}_{V1}$ and attempt to eliminate all instances of the field acceleration using combinations of the equations of motion. Inspired by the two-dimensional case, we first consider
\begin{align}
	\dot{\mathcal{C}}_{V1}+W\ud{i}{\mu} \p_i \mathcal{E}^{\mu} =& \p_i \ddot{A}_{\alpha} \left[ V_{\mu} \mathcal{H}^{i0,\mu\alpha} + W\ud{i}{\mu} \mathcal{H}^{\mu\alpha} \right] + \ddot{A}_{\alpha} \left[ V_{\mu} \p_i \mathcal{H}^{i0,\mu\alpha} + W\ud{i}{\mu} \p_i \mathcal{H}^{\mu\alpha} \right] \nn \\
	&+ V_{\mu} \p_i \left( \mathcal{H}^{ij,\mu \alpha} \p_j \dot{A}_{\alpha} \right) + W\ud{i}{\mu} \p_i \left( \mathcal{H}^{0j,\mu \alpha} \p_j \dot{A}_{\alpha} + \tilde{u}^{\mu} \right) \\
	&- 2 \Lambda^2 \p_t \left(\alpha_{0,X} A^{\mu} \right) V_{\mu} + \frac{1}{2\Lambda^{D/2}} \tilde{u}_{\mu} \tilde{u}^{\mu} \,,\nn
\end{align}
and observe that the last two lines do not involve second time derivatives, while the coefficient of $\p_i \ddot{A}_{\alpha}$ is in fact zero since
\ba
	0 = \frac{\p \eta^{i0}}{\p \dot{A}_{\alpha}} = \frac{\p (W^{i\mu} V_{\mu})}{\p \dot{A}_{\alpha}} = \left( \frac{\p W^{i\mu}}{\p \dot{A}_{\alpha}} V_{\mu} + W\ud{i}{\mu} \frac{\p V^{\mu}}{\p \dot{A}_{\alpha}}  \right) 
	\quad \Rightarrow \quad V_{\mu} \mathcal{H}^{i0,\mu\alpha} + W\ud{i}{\mu} \mathcal{H}^{\mu\alpha} = 0 \label{eq:HVplusHW} \,,
\ea
using the above identities.\\

It remains to eliminate the terms proportional to $\ddot{A}_{\alpha}$, which can in principle be achieved by adding a linear combination $\Upsilon_{\mu} \mathcal{E}^{\mu}$ of the undifferentiated equation of motion. It simplifies the calculation to separate the term $\p_i W\ud{i}{\mu} \mathcal{E}^{\mu}$ from the unknown vector $\Upsilon_{\mu}$, i.e.\
\begin{align}
	\dot{\mathcal{C}}_{V1} +\p_i \left(W\ud{i}{\mu} \mathcal{E}^{\mu} \right) + \Upsilon_{\mu} \mathcal{E}^{\mu} &=  \ddot{A}_{\alpha} \left[ \Upsilon_{\mu} \mathcal{H}^{\mu\alpha} -(\p_i V_{\mu}) \mathcal{H}^{i0,\mu\alpha} \right] \nn \\
	&\quad + V_{\mu} \p_i \left( \mathcal{H}^{ij,\mu \alpha} \p_j \dot{A}_{\alpha} \right) + \p_i \left( W\ud{i}{\mu} u^{\mu} \right) \\
	&\quad - 2 \Lambda^2 \p_t \left(\alpha_{0,X} A^{\mu} \right) V_{\mu} + \frac{1}{2\Lambda^{D/2}} \tilde{u}_{\mu} \tilde{u}^{\mu} + \Upsilon_{\mu} u^{\mu} \,.\nn
\end{align}
From this last result, we see that a second constraint exists provided the equation
\begin{equation}
	\Upsilon_{\mu} \mathcal{H}^{\mu\alpha} = (\p_i V_{\mu}) \mathcal{H}^{i0,\mu\alpha} \,,
	\label{eq:defUpsilon}
\end{equation}
admits a solution for $\Upsilon_{\mu}$.

\paragraph{Two-dimensions:}
    Note that in $D=2$ dimensions, we have the simple relation $\p_1 V_{\mu} = - \Lambda \Phi_1 W^1_{\mu}$. Since from \eqref{eq:HW}, we have $\mathcal{H}^{10,\mu \alpha}W^1{}_\mu=0$,  the RHS of \eqref{eq:defUpsilon} cancels exactly and there is  no need for a vector $\Upsilon$ in two dimensions. This agrees with the explicit two-dimensional derivation performed previously.

\paragraph{Higher Dimensions:} More generically, i.e. in arbitrary dimensions, since the Hessian $\mathcal{H}^{\mu\alpha}$ is non-invertible, it is clear that the solution of the previous equation is degenerate and indeed, $\Upsilon_{\mu}$ is not unique. Naturally, for any $\Upsilon_{\mu}$ that solves \eqref{eq:defUpsilon}, the vector $\Upsilon_{\mu}+f V_{\mu}$ is also a solution for any function $f$. This is perfectly consistent as it simply encodes the fact that the constraint $\mathcal{C}_{V2}$ can be shifted by  $fV_{\mu}u^{\mu}=f\mathcal{C}_{V1}\approx0$.

With this in mind, it will be useful to separate our Hilbert space into the NEV direction and its $(D-1)$-normal plane to derive the generic solution of \eqref{eq:defUpsilon}. We use the eigenvectors of the Hessian $\mathcal{H}^{\mu\nu}$ to span over our full $D$-dimensional Hilbert space and choose the set of $D$ eigenvectors $\{\tilde{V}_\mu^{(\sigma)}\}_{\sigma=0,\cdots,D-1} =  \{V_\mu, V^{\perp\, (a)}_\mu\}_{a=1,\cdots,D-1}$ which forms a complete orthonormal basis, satisfying the following properties
\beq\bal
\label{eq:norm}
& \mathcal{H}^{\mu \alpha}  \tilde V^{(0)}_{\alpha}=  \mathcal{H}^{i0,\mu \alpha}  \tilde V^{(0)}_{\alpha} = 0\,,  \\
& \mathcal{H}^{\mu \alpha} \tilde V^{(a)}_\alpha =\lambda^{(a)}\tilde V^{(a)\mu }\ne 0 \,, \ \   \forall \ a=1,\cdots,D-1,\\
 \text{and normalized as} &\  \tilde V^{(\sigma)}_\mu \tilde V^{(\sigma')\mu}=\eta^{\sigma \sigma'}\,, \ \  \forall \ \ \sigma, \sigma'=0\cdots,D-1\,.
\eal\eeq
Expanded in this basis, we have $\Upsilon_\mu=\upsilon_0 V_\mu +\upsilon_a V^{\perp (a)}_\mu$, with the component $\upsilon_0$ being arbitrary as discussed earlier, and the coefficients $\upsilon_a$ given by
\ba
\label{eq:upsilona}
\upsilon_a=\frac{1}{\lambda^{(a)}} (\p_i V_{\mu}) \mathcal{H}^{i0,\mu\alpha} V^{\perp (a)}_\alpha\ \forall \ a=1,\cdots, D-1\,.
\ea
Since $\lambda^{(a)} \ne 0$ for any situation  continuously related to the trivial vacuum, there is no ambiguity in uniquely identifying each coefficient $\upsilon_a$. See Appendix~\ref{app:Upsilon} for the explicit proof that the vector  $\Upsilon_\mu = \upsilon_a V_\mu^{\perp (a)}$ satisfies the relation \eqref{eq:defUpsilon}.
  This proves the existence of the required vector $\Upsilon_\mu$ and hence the existence of a secondary constraint in arbitrary dimensions.

Perturbatively, $\Upsilon_\mu$ is given by (up to  $\upsilon_0 V_\mu$),
\begin{align}
	\Upsilon_{\mu} =& \frac{1}{8\Lambda^D} \delta_{\mu}^i (\p^j A_0 + \dot{A}^j) (\p_i F_{j0} - \p_j F_{i0} ) \nonumber \\
	&+ \frac{1}{32 \Lambda^{3D/2}} \delta_{\mu}^i \left[ 2 \left( 2 \dot{A}_0 \dot{A}^j + \dot{A}^k F\ud{j}{k} - \p^k A_0 B\ud{j}{k} \right) (\p_i F_{j0} -\p_j F_{i0} ) \right. \nonumber \\
	& + 2 \left( \p_i A^k B\ud{j}{0} - \p_i A^j B\ud{k}{0} \right) (\p_j F_{k0} -\p_k F_{j0} )  \\
	&\left. + B\ud{j}{0} \left( 2\p_i \dot{A}_0 F_{j0} - 2\p_j \dot{A}_0 F_{i0} + B\du{j}{k} \p_i F_{0k} + 2 F\ud{k}{j} \p_i \dot{A}_k + B\ud{k}{0} \p_j F_{ik} + B\du{i}{k} \p_k F_{j0} + 2 F\du{i}{k} \p_j \dot{A}_k \right) \right] \nonumber \\
	&+ \mathcal{O}\left( \frac{(\p \p A) (\p A)^3}{\Lambda^{2D}} \right) \nn \,.
\end{align}
Note that this expression trivially cancels in $D=2$ dimensions, as previously highlighted. \\

To summarize this analysis, we have therefore proven the presence of a second Lagrangian constraint in arbitrary dimensions,
\begin{equation}
	\mathcal{C}_{V2} \equiv V_{\mu} \p_i \left( \mathcal{H}^{ij,\mu \alpha} \p_j \dot{A}_{\alpha} \right) + \p_i \left( W\ud{i}{\mu} u^{\mu} \right) - 2 \Lambda^2 \p_t \left(\alpha_{0,X} A^{\mu} \right) V_{\mu} + \frac{1}{2\Lambda^{D/2}} \tilde{u}_{\mu} \tilde{u}^{\mu} + \Upsilon_{\mu} u^{\mu} \approx0 \,,
\end{equation}
with $\Upsilon_{\mu} u^{\mu}$  given in terms of the non-null eigenvectors $V^{\perp (a)}_\mu$ and related eigenvalues $\lambda^{(a)}\ne 0$ of the Hessian by
\ba
\Upsilon_{\mu} u^{\mu}= \sum_{a=1}^D \frac{1}{\lambda^{(a)}} (\p_i V_{\nu}) \mathcal{H}^{i0,\nu\alpha} V^{\perp (a)}_\alpha  V^{\perp (a)}_\mu u^{\mu} \,.
\ea

\subsection{Hamiltonian picture}

Establishing the existence of a pair of constraints in the Hamilton-Dirac analysis is very simple for the minimal model \eqref{eq:LEPND} and in fact, proceeds very analogously to the two-dimensional case studied above. Using the identity \eqref{eq:trace derivative identity} with $n=1$ we derive the canonical momenta,
\beq
p^{\mu}=-2\Lambda^{D/2} V^{\mu} \,,
\eeq
in terms of the NEV $V^{\mu}$. The normalization of the latter then immediately allows us to infer the primary constraint
\beq
\mathcal{C}_1\equiv p^{\mu}p_{\mu} + 4 \Lambda^D \approx0 \,.
\eeq
This constraint trivially commutes with itself (recall that we take $\alpha_1$ constant for simplicity, but the conclusion also easily follows if $\alpha_1$ is a generic function), thus proving the existence of a secondary constraint, since $\mathcal{C}_1$ cannot be first class (since it is already not first class at the linear level).
Deriving the full expression of the canonical Hamiltonian in closed form is technically more challenging but not required to ascertain the existence of a secondary constraint. To close the algebra, one should then in principle check whether a tertiary or further constraints could exist (which would occur if $\mathcal{C}_1$ and $\mathcal{C}_2$ commute), however the arguments given in Section \ref{sec:reviewEPNConstraint} are proof enough that, at least for generic parameters, the theory cannot have \textit{fewer} than $D-1$ degrees of freedom (and if did at some point for some solutions and choices of parameters, these would not be trustable as the theory would then be infinitely strongly coupled at those points). We therefore conclude that there must be precisely $D-1$ degrees of freedom given the two constraints we have inferred.

\section{Discussion and conclusions}
\label{sec:ConclConstraint}

We have performed a constraint analysis of the recently proposed (Extended) Proca-Nuevo theory, with the particular aim of establishing the existence of a pair of constraints responsible for removing the Ostrogradsky ghost mode and thus rendering the model consistent from the point of view of the degree of freedom count, i.e.\ that a massive spin-1 system must describe $D-1$ dynamical modes in $D$ spacetime dimensions.

We devoted the first part of the paper to explaining why this outcome had to be expected. We showed through several formal and physical arguments why local, Lorentz and parity invariant field theories cannot hold a half number of Lagrangian degrees of freedom, equivalently an odd-dimensional physical phase space. Although these arguments are not new, they are certainly worth being recollected and emphasized in view of the fact that opposing claims have been made in the literature.

Summarizing our results concerning the analysis of EPN theory, we proved in full detail the existence of two, and only two, constraints in the general two-dimensional model, using both Lagrangian and Hamiltonian approaches. In the latter case, by deriving the full constraint algebra we moreover established the second class nature of the constraints and the absence of tertiary and further constraints. The generalization to arbitrary dimensions proved to be technically challenging, but we successfully analyzed a minimal version of the theory, concluding again the existence of a pair of constraints.

In addition to the main result regarding the counting of degrees of freedom, we find it worthwhile to remark on the interesting structure of the constraints we found in the canonical formalism. In particular, for the models we studied, the constraints are non-linear in all phase space variables and thus do not appear to smoothly deform the structures found in GP theory or linear Proca. We think this motivates a revisiting of the Hamiltonian analysis of EPN in arbitrary dimensions, where it may shed light on the issues related to the coupling to gravity of the theory.

\section{Acknowledgments}
CdR is supported by  STFC Consolidated Grant ST/T000791/1 and a Simons Investigator award 690508. SGS acknowledges support from the NSFC Research Fund for International Scientists (Grant No.\ 12250410250). VP is funded by an Imperial College President's Fellowship.
LH is supported by funding from the European Research Council (ERC) under the European Unions Horizon 2020 research and innovation programme grant agreement No 801781 and by the Swiss National Science Foundation grant 179740. LH further acknowledges support from the Deutsche Forschungsgemeinschaft (DFG, German Research Foundation) under Germany’s Excellence Strategy EXC 2181/1 - 390900948 (the Heidelberg STRUCTURES Excellence Cluster).

\appendix

\section{EPN constraints in the Lagrangian formalism}
\label{app:EPNLagrangian}

The analytic form of the acceleration-free part $u^{\mu}_{\rm (EPN)}$ of the equations of motion Eq.~\eqref{eq:EL} for the massive vector field $A_{\mu}$ in the two-dimensional EPN theory is given by
\begin{equation}
\begin{aligned}
u_{\rm (EPN)}^0 =& \; \frac{\tilde\alpha_1}{N_+^3} \left( x_{+} y_{+} \left( 2 \dot{A}_0' - A_1'' \right) - x_{+}^2 A_0'' + \left( x_{+}^2 + y_{+}^2 \right) \dot{A}_1' \right)  \\
	& + 2\Lambda \left( \tilde\alpha_{1,X} \frac{y_{+}}{N_{+}} + \alpha_{2,X} \left( y_+ + y_- \right) \right) \left( A_1 A_1' - A_0 A_0' \right)  \\
	& - 2\Lambda \left( \tilde\alpha_{1,X} \frac{x_{+}}{N_{+}} + \alpha_{2,X} (x_{+}+x_{-}) \right) (A_1 \dot{A}_1 - A_0 \dot{A}_0) \\
	& + 2 \Lambda^2 \left( \tilde\alpha_{0,X} + \tilde\alpha_{1,X} N_{+} + \frac12 \alpha_{2,X} \left( N_{+}^2 -N_{-}^2 \right) \right) A_0  \\
	&+  \Lambda^2 d_{1,X} \left( 2(1+\Sigma)A_0 - (y_++y_-)A_1 \right) \,, 
\end{aligned}
\end{equation}
and
\begin{equation}
\begin{aligned}
	u_{\rm (EPN)}^1 =& \; \frac{\tilde\alpha_1}{N_{+}^3} \left( x_{+} y_{+} \left( 2  \dot{A}_1' -  A_0'' \right) - y_{+}^2 A_1'' + \left( x_{+}^2 + y_{+}^2 \right)  \dot{A}_0' \right) \\
	& + 2\Lambda \left(  \tilde\alpha_{1,X} \frac{x_{+}}{N_{+}} + \alpha_{2,X} \left( x_+ - x_- \right) \right) \left( A_1 A_1' - A_0 A_0' \right)  \\
	& - 2\Lambda \left(  \tilde\alpha_{1,X} \frac{y_{+}}{N_{+}} + \alpha_{2,X} (y_{+}-y_{-}) \right) (A_1 \dot{A}_1 - A_0 \dot{A}_0)  \\
	& -2 \Lambda^2 \left( \tilde\alpha_{0,X} + \tilde\alpha_{1,X} N_{+} + \frac12 \alpha_{2,X} \left( N_{+}^2 -N_{-}^2 \right) \right) A_1  \\
	&+ \Lambda^2 d_{1,X} \left( 2\Delta A_0 - (2+x_+ -x_-)A_1 \right) \,.
\end{aligned}
\end{equation}
Furthermore, we defined the following functions entering the constraint Eq.~\eqref{eq:secondclassEVb},

\begin{equation}
\begin{aligned}
	\phi_0 =& 2 \left( \bar{x}_{+} A_0 + \bar{y}_{+} A_1 \right) \,, \\
	\phi_1 =& 2 \left( \left( 1 + \Sigma \right) A_0 + \Delta A_1  \right) \,, \\
	\phi_2 =& 4 \left( \left( \bar{x}_{+} \Sigma - \bar{y}_{+} \Delta \right) A_0 - \left( \bar{x}_{+} \Delta + \bar{y}_{+} \Sigma \right) A_1  \right) \,, \\
	\Phi_1 =& \frac{1}{\Lambda} \left( \bar{x}_{+} \p_1 \bar{y}_{+} - \bar{y}_{+} \p_1 \bar{x}_{+}  \right) \,,
\end{aligned}
\end{equation}
where we have introduced the following notation
\begin{equation}
\label{eq:notations_Sigma_Nabla}
	\Sigma = \frac12 (x_+ + x_-) = 1 + \frac{A_1'}{\Lambda} \,, \qquad \Delta = \frac12 ( y_+ - y_-) = - \frac{A_0'}{\Lambda} \,.
\end{equation}

\section{EPN Secondary constraint in the Lagrangian picture}
\label{app:EPNSecConstraint}

The secondary second-class constraint for the EPN theory is obtained by imposing the time derivative of the primary constraint $\mathcal{C}_{V1}$ presented in Eq.~\eqref{eq:secondclassEVb} to vanish. Its expression reads,
\begin{align}
\frac{\mathcal{C}_{V2}}{\Lambda^3} =& \alpha_{2,XX} \left(\frac{\dot{X}}{\Lambda} \phi_2 + 4 \left(A_0 \Delta -A_1 (1-\Sigma )\right) \left(2 A_0 \gamma _1+2 A_1 \left(\gamma _2-N_+\right)\right) \right) \\
   &+ 2 \alpha_{2,X} \left\lbrace \left(N_-^2-N_+^2\right) \bar{x}_+ +2 A_0 N_+ \Phi _1+2 N_+ - 2 \left(\left(\psi _1-\psi_2\right){}^2-A_1^2 N_+^2\right) \frac{\alpha_{2,X}}{\tilde{\alpha}_1} \right. \nonumber \\
   &\qquad \left. + \left(A_1 N_+^2 \psi _2+2 \left(\psi _2-\psi _1\right) \psi _3\right) \frac{\tilde{\alpha}_{1,X}}{\tilde{\alpha }_1} + 2 \psi _2 \left(2 A_0 \gamma _1+A_1 \left(2 \gamma _2-N_+\right)\right) \frac{\tilde{\alpha }_{0,X}}{\tilde{\alpha}_1} \right. \nonumber \\
   & \qquad \left. + 2 \left[ -4 A_0^2 \gamma _1 \left(\gamma _1-\bar{y}_+\right)-2 A_1 A_0 \left(2 \gamma _1 \left(2 \gamma_2-N_+\right)+y_-\right) \right. \right. \nonumber \\
   &\qquad \qquad \left. \left. + A_1^2 \left(2 \bar{x}_+ \left(2 \gamma _2-N_+\right)-4 \gamma _2 \left(\gamma
   _2-N_+\right)\right) \right] \frac{d_{1,X}}{\tilde{\alpha }_1} \right\rbrace \nonumber \\
   &+\tilde{\alpha }_{1,XX} \left( \frac{\dot{X}}{\Lambda} \phi_1 -4 \Delta \frac{\dot{A}_1}{\Lambda} A_0^2 +2 A_1 A_0 \left(\left(2-x_-\right) y_--x_+ y_+\right) +2 A_1^2 (1-\Sigma ) \left(-x_-+x_++2\right) \right) \nonumber \\
   &+\tilde{\alpha }_{1,X} \left\lbrace N_+ \Phi _1 \phi_0 +N_-^2-N_+^2+4 - \left(2 \psi _1 \psi _3- A_1 N_+^2 \psi _2\right)
   \frac{d_{1,X}}{\tilde{\alpha }_1} -2 \psi _2 \psi _3 \frac{\tilde{\alpha }_{0,X}}{\tilde{\alpha}_1} \right. \nonumber \\
   &\qquad \left. \vphantom{\frac{d_1}{\tilde{\alpha}_1}} + 2  \left( A_0^2 \Delta \left(y_-+y_+\right)-A_1 A_0 \left(\left(2-x_-\right) y_--x_+ y_+\right) -A_1^2 (1-\Sigma )
   \left(-x_-+x_++2\right) \right)\frac{\tilde{\alpha }_{1,X}}{\tilde{\alpha }_1} \right\rbrace \nonumber \\
   &+ \Phi_1^2 \tilde{\alpha}_1 +\tilde{\alpha }_{0,XX} \left(\frac{\dot{X}}{\Lambda}\phi_0 -2 \psi _2 \left(A_0 \Delta -A_1
   (1-\Sigma )\right)\right) \nonumber \\
   &+ \tilde{\alpha }_{0,X} \left(2 \left(2\bar{x}_+-N_+\right) - 2 \psi_1 \psi_2 \frac{d_{1,X}}{\tilde{\alpha}_1} - \psi_2^2 \frac{\tilde{\alpha}_{0,X}}{\tilde{\alpha}_1} \right) \nonumber \\
   &+ d_{1,X} \left(\left(N_-^2-N_+^2+4\right) \bar{x}_+ +2 N_+ \Phi_0 A_0 -\frac{d_{1,X}}{\tilde{\alpha }_1}\left(\psi _1-A_1 N_+\right) \left(A_1 N_++\psi _1\right) \right) \nonumber \\
   &+ d_{1,XX} \left(\frac{\dot{X}}{\Lambda} \left(\phi_0+\frac{\phi _2}{2}\right) -2 \left(A_0
   \Delta -A_1 (1-\Sigma )\right) \left(A_1 N_++\psi _1\right)\right) \,, \nonumber
\end{align}
where we defined
\begin{equation}
\begin{aligned}
	\psi_1 =& (y_- \bar{x}_+ + (2+x_-) \bar{y}_+) A_0 + ((2-x_-) \bar{x}_+ - y_- \bar{y}_+) A_1 \,, \\
	\psi_2 =& 2(\bar{y}_+ A_0 + \bar{x}_+ A_1) \,, \\
	\psi_3 =& y_- A_0 + (2-x_-) A_1 \,,
\end{aligned}
\end{equation}
and
\begin{equation}
\begin{aligned}
	\gamma_1 =& \bar{x}_+ \Delta - \bar{y}_+ \Sigma \,, \\
	\gamma_2 =& \bar{x}_+ \Sigma - \bar{y}_+ \Delta \,.
\end{aligned}
\end{equation}
It is now easy to see that, as claimed in the main text, the secondary constraint $\mathcal{C}_{V2}$ does not involve  any acceleration term.

\section{Secondary Constraint in arbitrary dimensions}
\label{app:Upsilon}
To complete the proof of the existence of a secondary constraint for the minimal model in arbitrary dimensions, we explicitly show that the vector $\Upsilon_\mu$ defined as in \eqref{eq:upsilona} by
\ba
\label{eq:UpsilonSol}
\Upsilon_\mu= \upsilon_a V^{\perp (a)}_\mu
= \sum_{a=1}^{D-1} \frac{1}{\lambda^{(a)}} (\p_i V_{\mu}) \mathcal{H}^{i0,\mu\alpha} V^{\perp (a)}_\alpha V^{\perp (a)}_\mu\,,
\ea
satisfies the relation  \eqref{eq:defUpsilon}.
First, by construction, we clearly have
\ba
\label{eq:UpsilonH}
\Upsilon_\mu \mathcal{H}^{\mu\alpha}= \upsilon_a V^{\perp (a)}_\mu \mathcal{H}^{\mu\alpha}
= \sum_{a=1}^{D-1}\upsilon_a \lambda^{(a)} V^{\perp (a)}{}^\alpha
= \sum_{a=1}^{D-1}  (\p_i V_{\mu}) \mathcal{H}^{i0,\mu\beta} V^{\perp (a)}_\beta V^{\perp (a)}{}^\alpha\,.
\ea
Now  recalling that as argued below Eq.~\eqref{eq:defUpsilon}, the  $D$ eigenvectors $\{\tilde{V}_\mu^{(\sigma)}\}_{\sigma=0,\cdots,D-1} =  \{V_\mu, V^{\perp\, (a)}_\mu\}_{a=1,\cdots,D-1}$ form a complete orthonormal basis, satisfying various properties, notably (see \eqref{eq:HW2} and \eqref{eq:norm}),
\ba
  \mathcal{H}^{i0,\mu \beta}   V_{\beta} = 0  \quad {\rm and }\quad
 V^{\perp\, (a)}_\mu V^{\perp\, (b)}{}^\mu=\delta^{ab} \,, \ \forall \ a,b=1,\cdots,D-1,
\ea
we can hence expand any vector in that complete basis. In particular we can write the vector
\ba
T^\beta=(\p_i V_\mu){\mathcal{H}}^{i0,\mu \beta}=\tau_0 V^\beta+\tau_b V^{\perp\, (b)}{}^\beta\,,
\ea
where $\tau_0=0$ since $V_\beta$ is also a NEV of  $\mathcal{H}^{i0,\mu \beta}$ and hence $T^\beta V_\beta =0=\tau_0$. With this in mind, we can therefore expand the RHS of \eqref{eq:UpsilonH} as follows
\ba
\label{eq:UpsilonH2}
\Upsilon_\mu \mathcal{H}^{\mu\alpha}&=&\sum_{a=1}^{D-1}  T^\beta V^{\perp (a)}_\beta V^{\perp (a)}{}^\alpha \nn \\
&=&\sum_{a=1}^{D-1}\sum_{b=1}^{D-1} \tau_b   V^{\perp\, (b)}{}^\beta V^{\perp (a)}_\beta V^{\perp (a)}{}^\alpha \nn \\
&=&\sum_{a=1}^{D-1} \tau_a  V^{\perp (a)}{}^\alpha \equiv T^\alpha= (\p_i V_\mu){\mathcal{H}}^{i0,\mu \alpha}\,.
\ea
This concludes the proof that the vector $\Upsilon_\mu$ given in \eqref{eq:UpsilonSol} does indeed satisfy the relation \eqref{eq:upsilona}. Interestingly, this relies non-trivially on the fact that $V_{\nu}$ is a null eigenvector of both $\mathcal{H}^{i0,\mu\nu}$ and $\mathcal{H}^{00,\mu\nu}$.

\pagebreak

\bibliographystyle{JHEP}
\bibliography{references_constraints}

\end{document}